\documentclass[referee,a4paper,12pt,traditabstract]{jswsc} 

\usepackage{amsmath}
\usepackage{derivative}
\usepackage{graphicx}
\usepackage{txfonts}
\usepackage{subfigure}
\usepackage{epstopdf}
\usepackage[displaymath,mathlines]{lineno}
\usepackage[authoryear,round]{natbib}
\usepackage[backref]{hyperref}
\usepackage{url}
\usepackage[]{siunitx}

\usepackage{glossaries-extra}
\newabbreviation{iss}{ISS}{International Space Station}
\newabbreviation{gcr}{GCR}{galactic cosmic rays}
\newabbreviation{sep}{SEP}{solar energetic particles}
\newabbreviation{let}{LET}{linear energy transfer}
\newabbreviation{sipm}{SiPM}{silicon photomultiplier}
\newabbreviation{adu}{ADU}{Active Detection Unit}
\newabbreviation{mc}{MC}{Monte Carlo}
\newabbreviation{mip}{MIP}{minimum-ionizing particle}
\glsdisablehyper

\hypersetup{colorlinks=true,citecolor=cyan,urlcolor=cyan,linkcolor=blue}

\begin{document}


   \title{A Neural-Network Framework for Tracking and Identification of Cosmic-Ray Nuclei in the RadMap Telescope}

   \titlerunning{A Neural-Network Framework for the RadMap Telescope}

   \authorrunning{Meyer-Hetling et al.}

   \author{L. Meyer-Hetling
          \inst{1,2}
          \and
          M.J. Losekamm\inst{1,2,3}
          \and
          S. Paul\inst{1,2}
          \and
          T. Pöschl\inst{4}
          }

   \institute{Technical University of Munich, School of Natural Sciences, Garching, Germany\\
              \email{\href{mailto:luise.meyer-hetling@tum.de}{luise.meyer-hetling@tum.de}}
         \and
             Excellence Cluster ORIGINS, Garching, Germany
         \and
             Now at European Space Agency, Noordwijk, Netherlands
         \and
             European Organization for Nuclear Reserach (CERN), Geneva, Switzerland
             }


   \abstract{
   We present a neural-network framework designed to reconstruct the properties of cosmic-ray nuclei traversing the scintillating-fiber tracking calorimeter of the RadMap Telescope. Employing the \textsc{Geant4} simulation toolkit and a simplified model of the detector to generate training and test data, we achieve the spectroscopic capabilities required for an accurate determination of the biologically relevant dose that astronauts receive in space. We can reconstruct a particle's trajectory with an angular resolution of better than $1.4^\circ$ and achieve a charge separation of better than 95\% for nuclei with $Z\leq8$; specifically, we reach an accuracy of 99.8\% for hydrogen. The energy resolution is $<20\%$ for energies below \SI{1}{\GeV/n} and elements up to iron.  We also discuss the limitations of our detector, the reconstruction framework, and this feasibility study, as well as possible improvements.}

   \keywords{cosmic-ray nuclei --
                radiation monitoring --
                tracking calorimeter --
                neural networks
               }

   \maketitle

\section{Introduction}

Mitigating the effects of exposure to cosmic and solar radiation on the human body is one of the major challenges of future missions to the Moon, Mars, and other deep-space destinations \citep{Chancellor2014}. Detailed knowledge of the radiation environment in space---including its spectral, temporal, and spatial variations---is a prerequisite for developing medical and operational radiation-protection measures \citep{Montesinos2021}. It is also essential for the design of new spacecraft, habitats, surface vehicles, and spacesuits \citep{Durante2011,Barthel2019}. Though research in the field has intensified in the past two decades, we still lack detailed long-term data on the radiation environment beyond low Earth orbit. We also have an incomplete understanding of how the complex radiation field in space interacts with spacecraft shielding and the human organism \citep{Vozenin2024,Chancellor2018,Walsh2019}. We know, however, that exposure to it results in short- and long-term health risks, including cancer \citep{Chancellor2014,NASEMcancer,Guo2022} and cardiovascular diseases \citep{Delp2016,Rikhi2020}. Recent studies suggest that (prolonged) exposure may also permanently impair cognitive functions such as learning and memory \citep{Parihar2015,Klein2021,Alaghband2023}. Altogether, health risks due to radiation exposure may be one of the most limiting factors in determining the maximum permissible length of deep-space missions \citep{Cucinotta2015c}.

We are developing a new class of radiation monitors to help address the shortage of in-situ data on the space radiation environment \citep{Walsh2019,Fogtman2023}. The charged-particle detectors at the heart of these instruments are capable of differentiating between nuclei of varying biological effectiveness---which to leading order depends on their nuclear charge, $Z$---and can measure their energy. The first instrument to operationally demonstrate the underlying technologies is the RadMap Telescope \citep{Losekamm2021}, which so far has collected data on the radiation environment inside the \gls{iss} between April 2023 and January 2024 \citep{Losekamm2023a}.

In this article, we present a neural-network-based framework for analyzing the data generated by the RadMap Telescope. We explain the challenges of interpreting the measurements of our detector and how the architecture of the framework addresses them. The results we show are based on benchmarked simulation data and provide best-case limits on the performance we can expect to achieve. We also discuss the advantages and challenges of relying on deep-learning algorithms, as well as our approaches to tackle the latter. 

\section{The Space Radiation Environment}

The space radiation environment is a complex field of charged particles and atomic nuclei. It is dominated by two major components: \gls{gcr} and \gls{sep}. 98\% of \gls{gcr} are fully ionized nuclei of all naturally occurring elements, with protons (87\%) and helium nuclei (12\%) being most abundant; only about 1\% are nuclei of heavier elements \citep{Longair2012,Rankin2022}. The remaining 2\% are mostly electrons and positrons. \gls{gcr} nuclei have energies between a few \unit{\MeV} per nucleon (\unit{\MeV/n}) and hundreds of \unit{EeV}. Most relevant to radiation protection is the \unit{\MeV}-to-\unit{\GeV} range---around the spectrum's peak at about \qty{300}{\MeV/n}. Though the flux of protons is orders of magnitude larger than that of heavier nuclei, the latter's higher biological effectiveness results in them contributing more than half of the radiation dose astronauts receive \citep{Naito2022}. 

\gls{sep} are protons and light nuclei with energies up to several \unit{\GeV} that are released by the Sun in sudden energetic outflows of coronal matter \citep{Reames2021,Reames2022}. Such \gls{sep} bursts deliver high, quasi-instantaneous doses that can cause acute radiation sickness. In lightly shielded environments, the strongest events may even be fatal \citep{Hellweg2020,Mao2021}. Understanding the temporal and spatial evolution of \gls{sep} bursts is thus crucial for establishing operational procedures to limit the crew's exposure. Close to Earth and other planets with a magnetic field, particles (mostly protons and electrons) trapped in planetary radiation belts constitute a third major source of radiation exposure.

Uncertainties in the assessment and prediction of the risk posed by exposure to the space radiation environment arise mainly from our limited knowledge of the biological effectiveness of cosmic-ray nuclei \citep{Cucinotta2017,Chancellor2018}. Another source of uncertainty is that most instruments used for operational radiation monitoring cannot resolve the charge, $Z$, and (kinetic) energy, $E_\mathrm{kin}$, of a particle or nucleus (we use the terms interchangeably throughout this article) but instead measure only the \gls{let} or, worse, the energy deposition in a planar detector (e.g., a silicon diode). Though the composition of the radiation field can to some degree be determined from a recorded \gls{let} or energy-deposition spectrum when using certain model assumptions, this reconstruction suffers from sizeable uncertainties. The accurate determination of the biologically relevant dose therefore requires the development of new detector systems with spectroscopic capabilities \citep{Cucinotta2015}.

\begin{figure}
    \centering
    \includegraphics[width=.7\linewidth]{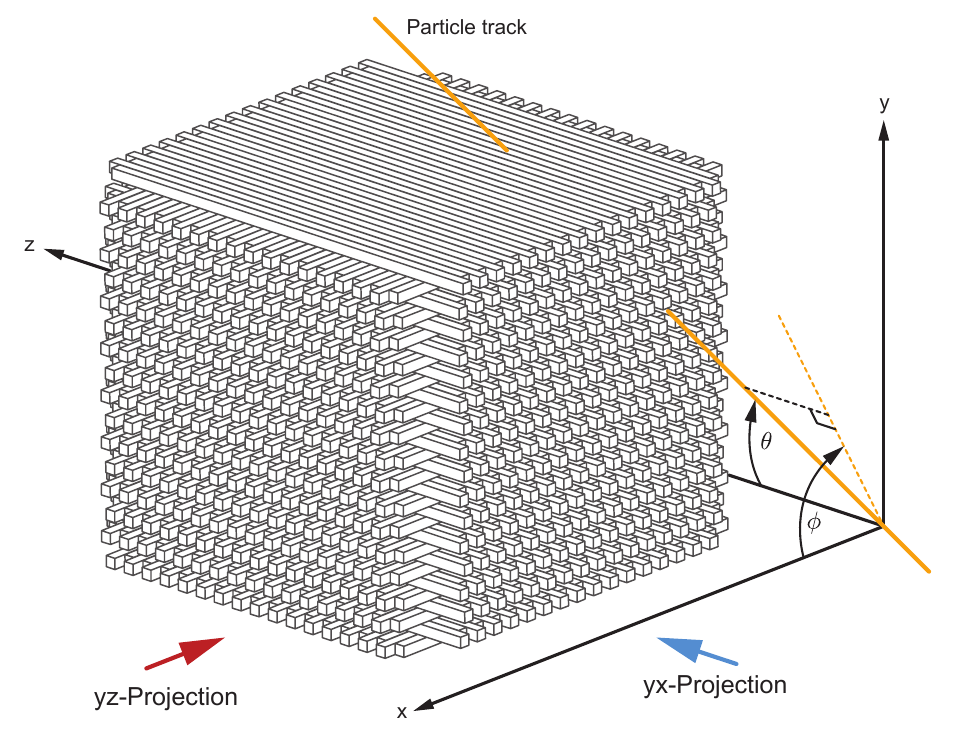}
    \caption{Schematic representation of the RadMap Telescope's main detector, which consists of a stack of 1024 scintillating plastic fibers. They are arranged in a stack of 32 layers of alternating orientation. Also shown are the coordinate system and the spherical coordinates $\phi$ and $\theta$ we use to parametrize the orientation of particle tracks. The red and blue arrows indicate the two-dimensional projections we use to visualize and analyze the detector's data.}
    \label{fig:adu-projections}
\end{figure}

\section{Measurement Principle and Reconstruction Methodology}
\label{Sec:Principle}

Instruments employed for the continuous characterization of the radiation environment that astronauts are exposed to must be small and consume little power. In contrast to those built for astrophysical investigations, they are hence often limited in the choice of technologies and in the complexity of their layout. The use of magnetic spectrometers for momentum measurement and charge determination, for example, is often not possible due to mass and power constraints. Other limitations arise because of environmental conditions or safety considerations.

\subsection{Detector Geometry}
\label{sec:Geometry}

We are attempting to overcome the limitations of many current-generation detectors by using a novel detection principle. Figure~\ref{fig:adu-projections} shows a schematic representation of the RadMap Telescope's main detector, the \gls{adu}. Its radiation-sensitive volume consists of 1024 scintillating-plastic fibers with a square cross-section of $\qtyproduct{2 x 2}{\mm}$ and a length of \qty{80}{\mm}. They are arranged in a stack of 32 layers of alternating orientation. The intensity of the scintillation light created in each fiber is, to leading order, proportional to a particle's energy loss in it. We detect the light using \glspl{sipm} attached to one end of each fiber. This layout allows recording the total energy deposition of individual particles and the energy-deposition profile along their tracks.

\begin{figure}
    \centering
    \includegraphics[width=.8\textwidth]{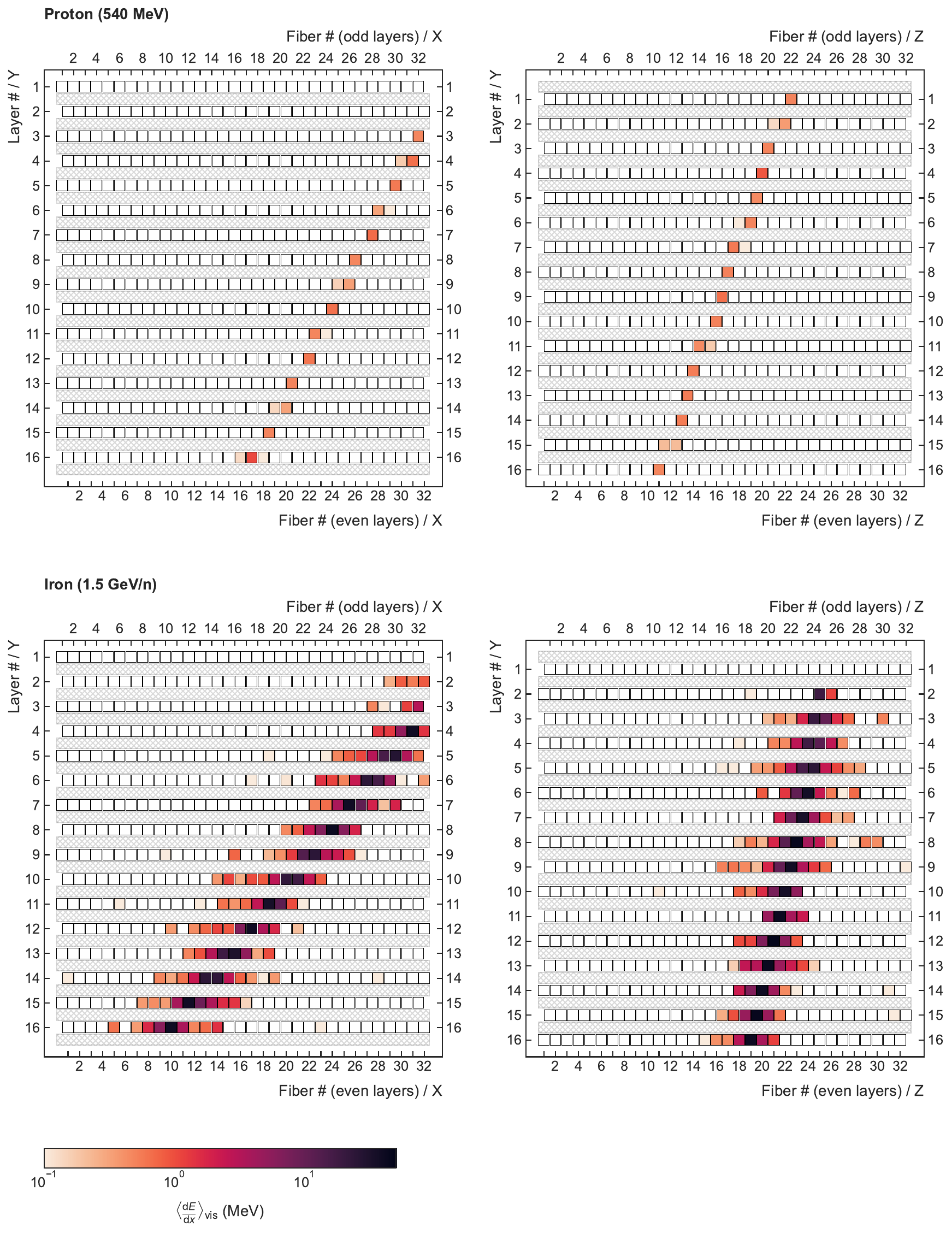}
    \caption{Simulated event signatures of a 540-\unit{MeV} proton (top) and a 1.5-\unit{GeV/n} iron nucleus (bottom), shown in the yx- (left) and yz-projections (right) illustrated in Figure~\ref{fig:adu-projections}. The color indicates how much energy is converted into scintillation light in each fiber.}
    \label{fig:detector-images}
\end{figure}

Figure ~\ref{fig:adu-projections} also shows the coordinate system and the spherical coordinates $\phi \in [-180^\circ,180^\circ)$ and $\theta \in [0^\circ,180^\circ]$ we use to parametrize the orientation and direction of particle tracks. The coordinate system's origin is at the center of the detector's x- and z-dimensions. We project the signal amplitudes along the two fiber orientations onto the yx- and yz-planes (see red and blue arrows in the figure), obtaining two gray-scale images with $16 \times 32$ pixels each. For illustration, Figure~\ref{fig:detector-images} shows the simulated event signatures (see Section~\ref{sec:training-data} below for a description of our \textsc{Geant4} setup) of a 540-\unit{MeV} proton and a 1.5-\unit{GeV/n} iron nucleus. The latter is much broader due to the emission of high-energy $\delta$ electrons and hadronic fragments along the track.

\subsection{Measurement Principle and Challenges}
\begin{figure}
    \centering
    \includegraphics[width=0.7\linewidth]{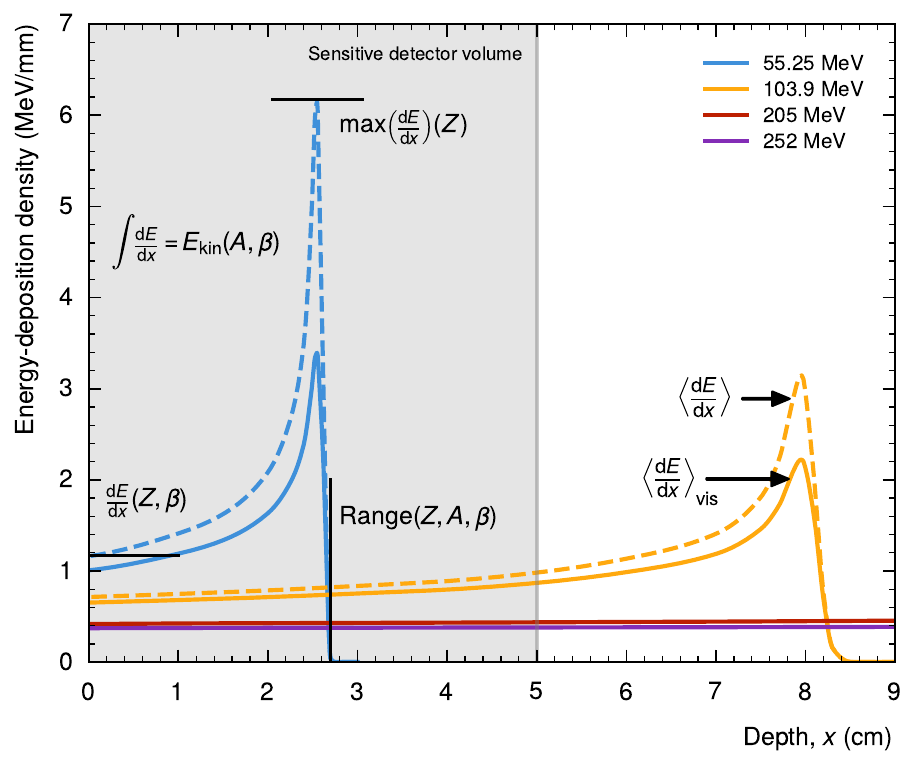}
    \caption{Illustration of the parameters that can be determined from the energy-deposition profiles of cosmic-ray nuclei. The dashed curves show the mean energy deposition, $\langle \odv{E}{x} \rangle$, of protons with different energies in high-density polyethylene$^{\mathrm{a}}$. The energy-deposition profile of a stopping particle encodes its nuclear charge, $Z$, mass number, $A$, velocity, $\beta$, and kinetic energy, $E_\mathrm{kin}$. For through-going particles, only $Z$ and $\beta$ can be determined. The solid curves illustrate how ionization quenching reduces the measurable energy-deposition density, which we denote as $\langle \odv{E}{x} \rangle_\mathrm{vis}$. For clarity, fluctuations due to energy-loss straggling are not shown. \\
    $^{\mathrm{a}}${\scriptsize\url{https://www.bnl.gov/nsrl/userguide/bragg-curves-and-peaks.php}}}
    \label{fig:bragg}
\end{figure}

The energy-deposition profile of cosmic-ray nuclei stopping in the detector contains all information required for determining their identity and energy. Gruhn et al.~first described the corresponding reconstruction technique, which they called \emph{Bragg curve spectroscopy} \citep{Gruhn1982}. Figure~\ref{fig:bragg} shows how a single measurement of a Bragg curve in an ideal detector (dashed blue curve) yields information about a particle's nuclear charge, $Z$, mass number, $A$, and kinetic energy, $E_\mathrm{kin}$ (the latter being a function of $A$ and the velocity, $\beta$). Since there is little difference in the biological effectiveness of isotopes of the same element, only $Z$ and $E_\mathrm{kin}$ are relevant for radiation dosimetry. If we assume that cosmic-ray nuclei are always fully ionized and that their isotopic composition is known and constant over time, $E_\mathrm{kin}$ can be inferred from $Z$ and $\beta$ using the average atomic mass. Thus, even for particles passing through the detector (see orange, red, and purple curves in Figure~\ref{fig:bragg}), the charge and velocity dependence of the energy loss contains sufficient information for identifying nuclei and measuring their energy. At higher energies, however, the resolution intrinsically decreases because the velocity dependence becomes less pronounced (see red and purple curves).

There are three effects that complicate this simplified picture. First, the statistical nature of the electronic interactions between nuclei and the detector material causes fluctuations of the formers' energy loss (called \emph{straggling}). For clarity, we show in Figure~\ref{fig:bragg} only the mean energy deposition, $\langle \odv{E}{x} \rangle$, of an ensemble of particles; the profiles for individual particles of the same charge and energy can differ significantly from this average. Second, the spallation of incident nuclei into fragments with lower $Z$ (and consequently smaller energy loss) leads to further variation in the energy-deposition profiles. Nuclei may fragment immediately upon entering the detector, somewhere along their path through it, or not at all. Our particle-identification and energy-reconstruction algorithms must therefore be able to cope with a broad range of potential energy-loss profiles for nuclei with identical $Z$ and $E_\mathrm{kin}$. This effect becomes stronger with increasing $Z$. Third, the scintillation efficiency---i.e., the fraction of energy deposited in a fiber that is converted into detectable light (see solid curves in Fig.~\ref{fig:bragg})---decreases with increasing energy-deposition density due to ionization quenching \citep{Poschl2020}. This effectively results in an upper limit on the measurable energy-deposition density, reducing the separation power for heavier nuclei because of the quadratic $Z$-dependence of the energy loss.

\subsection{Reconstruction Methodology}

\begin{figure}
    \centering
    \includegraphics[width=.9\textwidth]{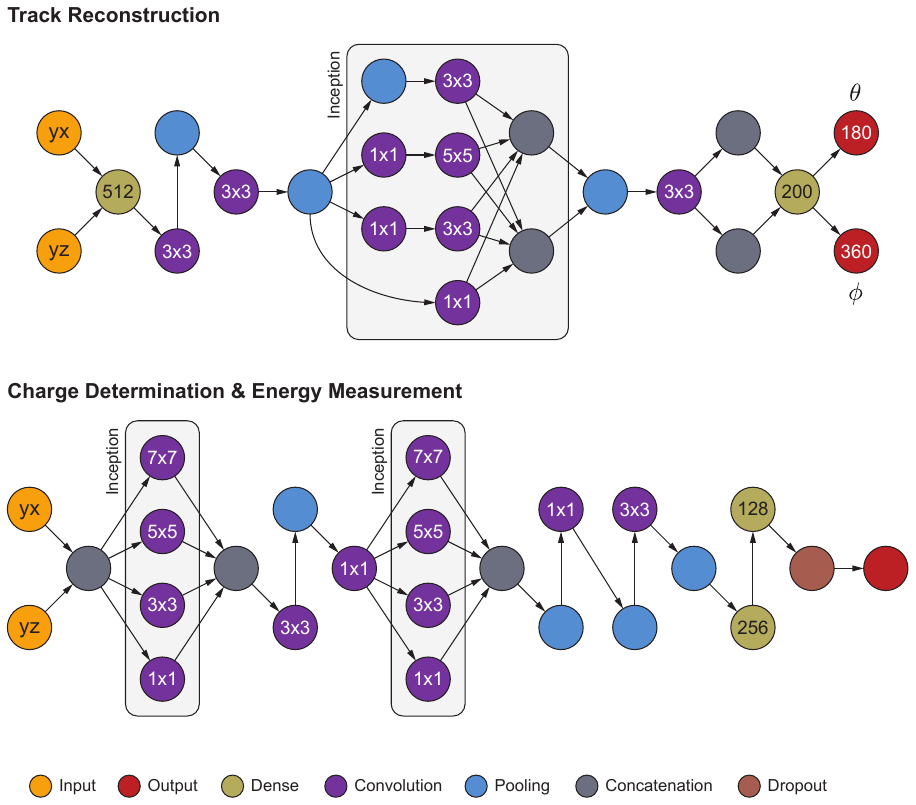}
    \caption{Architectures of the neural networks we use. Each circle represents a network layer and the color indicates the layer type. Dense and convolutional layers are labeled with their total size and their filter size, respectively. The network for track reconstruction has about 2.8 million trainable parameters, the one for charge determination and energy measurement about 2.1 million.}
    \label{fig:NN_architectures}
\end{figure}

Our goal is to perform an event-by-event analysis of the data gathered by the RadMap Telescope, i.e., to determine the properties of each individual particle detected by the \gls{adu}. For this, we need to (1) reconstruct the track of the particle through the detector, (2) determine its charge, and (3) calculate its initial kinetic energy. In the past, we tested multiple approaches for an automated event reconstruction, including a Bayesian particle filter \citep{Losekamm2017}, simulated annealing \citep{Laarhoven1987}, and a Markov Chain Monte Carlo \citep{Caldwell2009}. Though some of these methods produced acceptable results in a limited parameter space \citep{Hollender2019,Poschl2022}, their biggest drawback was the computing effort required, which resulted in processing times upwards of 15 minutes per event \citep{Milde2016,Losekamm2017}. Since our objective is to build a system that can operate in real time, such single-particle execution times are orders of magnitude too long.

We thus developed an alternative approach for event reconstruction based on neural networks that we train on simulated interactions of cosmic-ray nuclei with the \gls{adu}. Because each of the reconstruction steps depends on the result of the previous one, we developed three sequentially-applied frameworks. For each of them, we use the unaltered projections of the event signatures (see Fig.~\ref{fig:detector-images}) as input.

The networks at the heart of these frameworks have architectures with a similar basic structure (see Fig.~\ref{fig:NN_architectures}). Their main building blocks are convolutional layers, a layer type designed specifically for recognizing patterns in image-like data \citep{Lecun1998}. Convolutions divide the input images into smaller areas and attempt to extract local features---like edges or endpoints---before merging the areas back together to find a global structure---e.g., a straight line. They are well suited for detecting translation- and rotation-invariant patterns, for example the straight lines of particle tracks or the energy-deposition profiles along such tracks. The size of the areas---the filter size---is a key parameter of a convolutional layer, since it determines the initial scale at which the network searches for features. The scale of the track left by a particle in our detector, however, cannot be generally defined because it depends, among other things, on its orientation, its entrance point, and on the particle's properties. We thus rely on inception layers that combine multiple parallel convolutions with different filter sizes \citep{Szegedy2015}. This architecture allows the networks to learn by themselves which initial scale is best suited to finding the patterns we are looking for.

\subsection{Training Data}
\label{sec:training-data}
We generate the data used for the training, validation, and testing of the neural networks with the \textsc{Geant4} \gls{mc} simulation toolkit (version 11.2) using its standard physics list, FTFP\_BERT \citep{Agostinelli2003,Allison2016}. We implemented detector-specific effects like ionization quenching for a realistic representation of the physical processes leading to signal formation in the \gls{adu}. For the purposes of this article, the detector model consisted of only the stack of 1024 scintillating fibers. We deliberately did not include any material around the detector because it is a significant source of scattering and fragmentation, and we wanted to assess the \gls{adu}'s performance in the most unbiased way possible. In the extreme case of thick shielding---as provided by the \gls{iss}, for example \citep{Koontz2005}---the fragmentation probability for heavy nuclei is nearly 100\%, which means they can almost never be observed inside the spacecraft \citep{Zeitlin2016}. Recreating this environment would not have allowed us to study the reconstruction performance for all but the lightest elements.

We concentrated on the \gls{gcr} component of the space radiation environment and chose particle and energy distributions that ensured an unbiased training of the networks. We simulated the \gls{gcr} spectrum using the most abundant isotopes of all naturally occurring elements from hydrogen to iron (i.e., $^1$H, $^4$He, $^7$Li, ..., $^{58}$Fe). We did not include electrons and gamma rays because they are not relevant to our work. Furthermore, we assumed all elements to be equally abundant to prevent the networks from being biased towards certain elements. 

Likewise, we did not use a fully realistic power-law spectrum for modeling the \gls{gcr} energy distribution to avoid introducing a significant bias towards lower energies. Instead, we used a log-uniform distribution with task-specific energy ranges, which does take into account that smaller energies are more probable but still ensures that our training data includes a sufficiently large number of nuclei with higher energies. 
The nuclei were created on a spherical surface surrounding the detector, with an angular distribution that follows a cosine law to make the incident flux isotropic. Unless stated otherwise, we included only events for which a particle was detected in at least three fibers of each orientation ($N_{\rm sig}^{yx} \geq 3$ and $N_{\rm sig}^{yz} \geq 3$).

\section{Track Reconstruction}

Full track reconstruction requires determining the angles $\theta$ and $\phi$ (see Fig.~\ref{fig:adu-projections}) and the position of the track. We are, however, primarily interested in the two angles because they are all that is required to determine the angular distribution of the incident particle flux. $\phi$ can be determined directly from the yx-projection; $\theta$ must be calculated from its projection onto the yz-plane, $\theta_\mathrm{proj}$, via 
\begin{equation}\label{Eq:theta}
\theta = \arctan\Big(\frac{\tan\theta_\mathrm{proj}}{\sin\phi}\Big).
\end{equation}
Note that $\theta \in [0^\circ,180^\circ]$ and that we choose the value range of the arctan accordingly.

The neural network for track reconstruction has two output layers, one for $\theta_\mathrm{proj}$ and one for $\phi$ (see Fig.~\ref{fig:NN_architectures}). It performs a dual classification task with a bin width of $0.2^\circ$. Although using a regression approach seemed more intuitive to determine the two real-valued angles at first, a classification yielded significantly better results with an architecture of comparable complexity.  The network's output therefore consists of two discrete pseudo-probability distributions over 900 and 1800 classes, respectively. The reconstructed value for each angle is given by the output class with the highest attributed probability.

We trained the network on 0.7 million simulated events, reading the full data set in each epoch. We used the AdamW optimizer \citep{Loshchilov2019} to update the network weights and selected the final model via early stopping to avoid overfitting \citep{Prechtelt1998}: After each epoch, the performance of the network was validated on a distinct data set of 0.2 million events, and the result of each training iteration was saved only if the reconstruction performance improved with respect to the last result.

For track reconstruction, we require a more restrictive event selection than outlined above, with $N_{\rm sig}^{yx} \geq 5$ and $N_{\rm sig}^{yz} \geq 5$. In addition, we selected only particles that reached the detector's central region, a cuboid measuring (24 layers $\times$ 24 fibers $\times$ 24 fibers).

\subsection{Results}

\begin{figure}
    \centering
    \includegraphics[width=.77\linewidth]{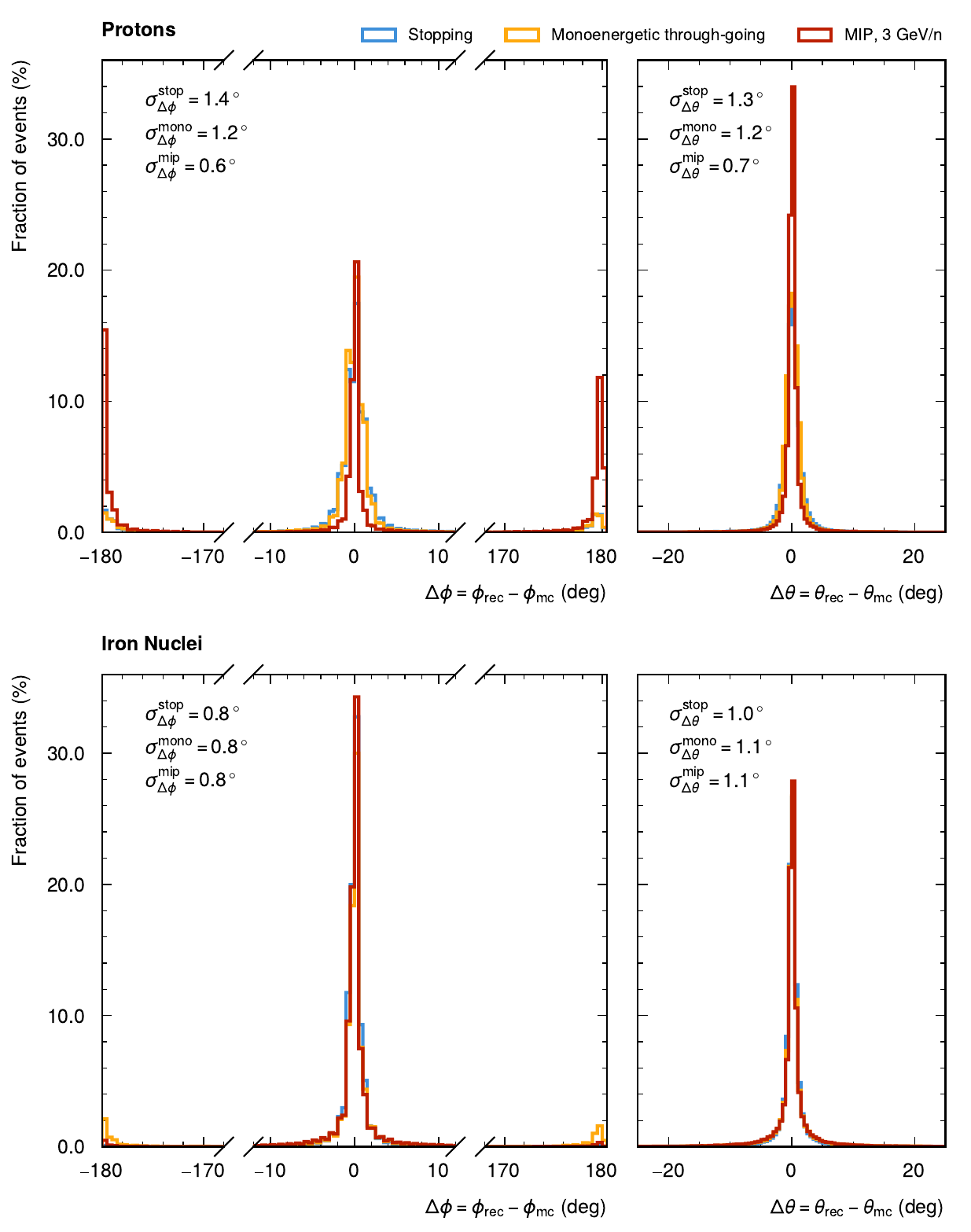}
    \caption{Reconstruction performance for the track angles $\phi$ (left) and $\theta$ (right) for protons and for iron nuclei,  shown as the differences ($\Delta\phi$ and $\Delta\theta$) between the reconstructed angles ($\theta_{\rm rec}$ and $\phi_{\rm rec}$) and the true angles ($\theta_\mathrm{mc}$ and $\phi_\mathrm{mc}$). The blue histograms show the performance using networks trained to reconstruct the tracks of stopping particles. The yellow histograms show the corresponding results for monoenergetic particles (\SI{120}{\MeV} for protons and \SI{500}{\MeV/n} for iron), the red histograms for minimum-ionizing particles with an energy of \SI{3}{GeV/n}. For each case, we quote the $\sigma$ of the direction-independent $\Delta\phi$ and $\Delta\theta$ distributions.}
    \label{fig:AngleResolution}
\end{figure}

We assessed the capabilities of the network architecture by training it on three different data sets each for both protons and iron nuclei. The first consisted of particles that stopped in the detector, which we manually selected using knowledge from the \textsc{Geant4} simulation. The second comprised monoenergetic, through-going particles with energies of \SI{120}{\MeV} for protons and \SI{500}{\MeV/n} for iron nuclei. This ensured that the particles' energy-deposition density increased significantly along their track (see yellow curve in Fig.~\ref{fig:bragg}). Finally, the third data set contained \glsxtrfullpl{mip} with an energy of \SI{3}{\GeV/n}. Each data set contained 0.1 million events. 

As figure of merit we use the differences between the reconstructed angles ($\theta_{\rm rec}$ and $\phi_{\rm rec}$) and the true angles from the \textsc{Geant4} simulation ($\theta_\mathrm{mc}$ and $\phi_\mathrm{mc}$), i.e., $\Delta \theta = \theta_{\rm rec} - \theta_\mathrm{mc}$ and $\Delta \phi = \phi_{\rm rec} - \phi_\mathrm{mc}$. Figure~\ref{fig:AngleResolution} shows the resulting distributions of $\Delta\theta$ and $\Delta\phi$. They are well centered around zero for all data sets, exhibiting only minimal shifts, $\mu$, of their mean values: $|\mu_{\Delta\phi}| \leq 0.07^\circ$ and $|\mu_{\Delta\theta}| \leq 0.1^\circ$ for protons and $|\mu_{\Delta\phi}| \leq 0.08^\circ$ and $|\mu_{\Delta\theta}| \leq 0.12^\circ$ for iron nuclei. All $\mu$ are smaller than the bin width of $0.2^\circ$, showing that our reconstruction is not significantly biased.

In the distribution of $\Delta\phi$ for minimum-ionizing protons, we observe that about 28\% of events are reconstructed with $\left|\Delta\phi\right| \approx 180^\circ$. That is because the network is able to correctly determine the orientation of these tracks but not the direction in which the particle is traveling. For nuclei whose energy-deposition density changes very little throughout the detector, the network does not have sufficient information to correctly determine the direction. Consequently, the fraction of events with an inverted reconstructed direction is significantly reduced for 120-MeV and for stopping protons. In the former case, the energy deposition changes substantially along their track while in the latter case it exhibits a Bragg peak and the corresponding track has an entry but no exit point.

To account for our inability to reconstruct the track direction for minimum-ionizing protons, we examined the distributions of $\Delta\phi$ and $\Delta\theta$ for direction-independent tracks. The definition of the track orientation is then ambiguous, with two combinations of $\phi$ and $\theta$ describing the same track. The two allowed values for $\phi$ are off by $180^\circ$. According to Equation~\ref{Eq:theta}, this translates into two possible values for the other angle, $\theta$ and $180^\circ-\theta$. To account for this ambiguity, we manually determined the track's direction based on our knowledge from the \gls{mc} data. This allowed us to select the correct $(\phi,\theta)$ combination for the computation of $\Delta\phi$ and $\Delta\theta$. We assume their distributions to be Gaussian and centered at zero, and define their standard deviation, $\sigma$, via the 68\% confidence interval.

\begin{figure}
\centering
		\includegraphics[width=.9\linewidth]{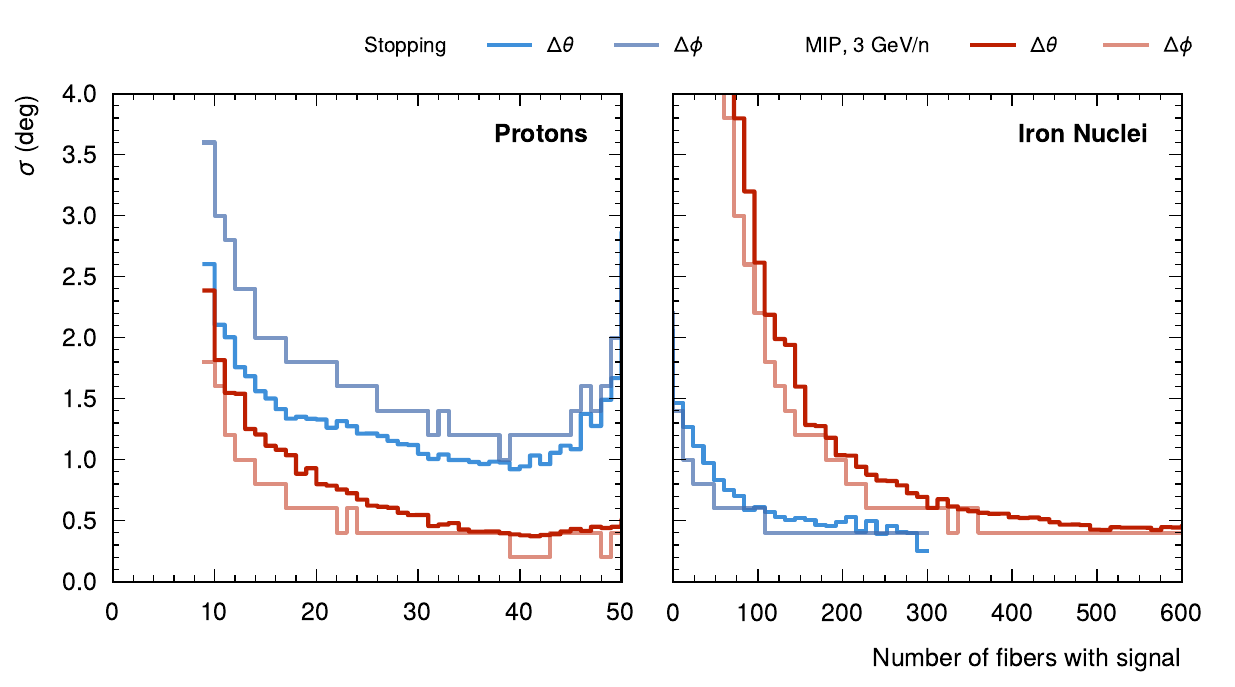}
		\caption{Dependence of $\sigma$ on the number of fibers with signal per event for stopping particles (blue) and minimum-ionizing particles (red).  Values of $\sigma_{\Delta\phi}$ are multiples of the neural network's output resolution ($0.2^\circ$); $\sigma_{\Delta\theta}$ is calculated via Equation~\ref{Eq:theta}. The number of traversed fibers can be larger than the depth of the \gls{adu} (32 fibers/layers) due to the emission of secondary particles (see Fig.~\ref{fig:detector-images}).}
		\label{fig:AnglesVsHitNumber}
\end{figure}

For minimum-ionizing protons, we achieve $\sigma^\mathrm{mip}_{\Delta\phi} = 0.6^\circ$ and $\sigma^\mathrm{mip}_{\Delta\theta} = 0.7^\circ$. This demonstrates that the network has learned correctly that adjacent output classes correspond to small angular differences for both angles. For stopping ($\sigma^\mathrm{stop}_{\Delta\phi} = 1.4^\circ$ and $\sigma^\mathrm{stop}_{\Delta\theta} =1.3^\circ$) and monoenergetic protons ($\sigma^\mathrm{mono}_{\Delta\phi} = 1.2^\circ$ and $\sigma^\mathrm{mono}_{\Delta\theta} = 1.2^\circ$), we obtain slightly larger values. The event signatures of protons with lower energies on average contain a smaller number of fibers with signal, making the track reconstruction more difficult for the network (or any other algorithm). This is shown in Figure~\ref{fig:AnglesVsHitNumber}.

In contrast, the reconstruction performance for iron nuclei does not exhibit a notable energy dependence and we obtain very similar values for all three examined cases: $\sigma_{\Delta\phi} = 0.8^\circ$ and $\sigma_{\Delta\theta} \leq 1.1^\circ$. This is because iron nuclei create a broader energy-deposition signature than protons (see Fig.~\ref{fig:detector-images}). The resulting higher information content of the events makes it easier for the network to reconstruct the particle tracks. Notably, it has significantly less difficulty in determining the direction because the track structures become wider along the direction of motion.  

\subsection{Discussion}
The angular resolutions of our track reconstruction for protons and iron nuclei ($\sigma_{\Delta\phi}\leq 1.4^\circ$ and $\sigma_{\Delta\theta} \leq 1.3^\circ$) are fully in line with the requirements of radiation monitoring. Using a simple geometric approximation, we estimated that the performance of our network is in fact close to the ideal limit imposed by the detector's effective pixel size of $2 \times 2\,\si{mm^2}$ \citep{Poschl2022}. 
Nonetheless, the values must be interpreted as a limit on the achievable resolution. In the real detector, deviations in the placement of the scintillating fibers lead to reconstruction uncertainties. Other effects, for example a non-uniform light yield of the fibers and variations in the \gls{sipm} response, contribute as well. The study of these uncertainties remains future work and is beyond the scope of this paper.

\section{Charge Determination}
\label{Sec: PID}

The \gls{adu} can identify nuclei only via their specific energy loss, which is a function of their charge and velocity (see Fig.~\ref{fig:bragg}). For stopping particles, the energy-deposition profile (whose reconstruction requires knowledge of the track parameters) would in principle also allow to determine a nucleus' mass number, $A$. We are, however, not interested in $A$ because it (a) is largely irrelevant for radiation dosimetry and (b) does not have a noticeable effect (within the resolution of our detector) for all but the lightest nuclei. In the simplified case of an unshielded detector that is subjected to a modified \gls{gcr} environment as described in Section~\ref{sec:training-data}, we can also assume that all primary particles with $\left|Z\right| = 1$ are protons. Other relevant particles with this charge (e.g., deuterons, muons, pions, electrons, and positrons) are only created as secondaries in the detector volume. In the context of the work presented here, determining the charge of particles therefore allows us to unambiguously identify nuclei of all elements from hydrogen to iron.

While the neural-network architecture for track reconstruction contains only a single inception layer of stacked convolutions, we found that using multiple flat inception layers is advantageous for charge determination. The number of classes in the output layer corresponds to the range of $Z$ we attempt to identify. For each event, the network computes a pseudo-probability distribution over all possible $Z$, and we select the class with the highest attributed probability as the reconstructed value.

We initially attempted to identify all $Z$ with a single network, achieving acceptable results for light elements but poor performance for heavier ones. This is plausible because the effects leading to deviations from the ideal energy-deposition profile (straggling, fragmentation, and quenching) become stronger for increasing $Z$. We therefore perform a two-step charge determination: A first network attempts to identify nuclei from hydrogen ($Z = 1$) through oxygen ($Z = 8$), sorting all particles it believes to be of higher charge into an additional `overflow' class. The events in this class are passed on to a second network that tries to identify fluorine ($Z = 9$) through iron ($Z = 26$). This approach greatly improves the reconstruction performance. We used the same training methodology as for the track reconstruction network (with nine million training events and one million testing events) but included cobalt ($Z = 27$) in the data set to avoid boundary effects for iron. The nuclei had energies between \SI{20}{MeV} and \SI{5}{TeV}.

\subsection{Results}
\begin{figure}[t]
\centering
		\includegraphics[width=0.7\linewidth]{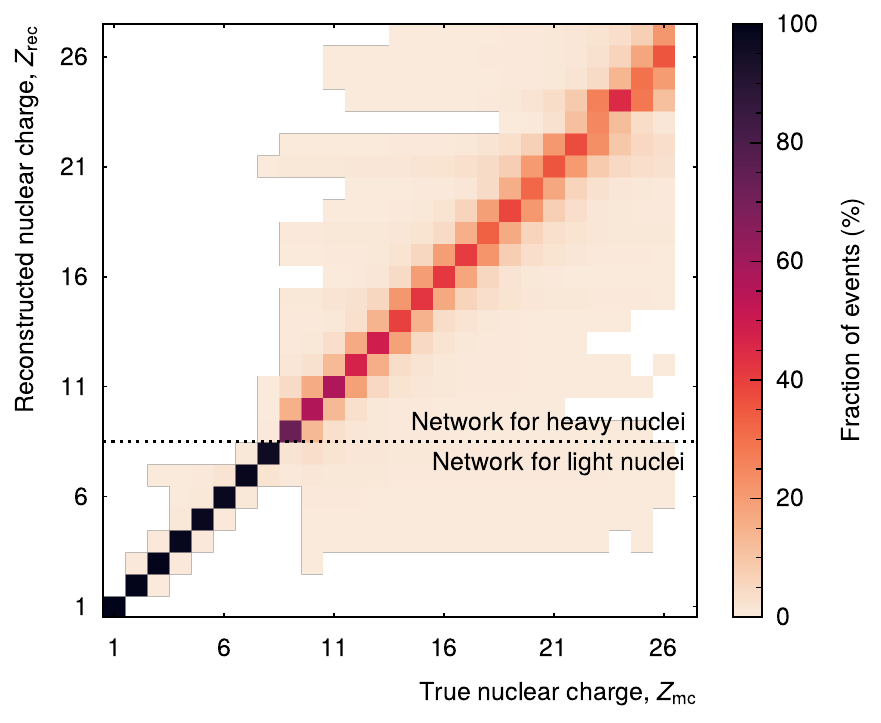}
        \caption{Confusion matrix of the charge determination via the consecutive application of two neural networks, trained to identify light nuclei through oxygen ($Z = 8$) and heavy nuclei through iron ($Z = 26)$, respectively. The second network was trained (but not tested) on a data set containing cobalt ($Z = 27$). Each column shows the distribution of reconstructed nuclear charge, $Z_{\rm rec}$, for all events with true nuclear charge, $Z_\mathrm{mc}$. Cells with values smaller than 0.1\% are drawn in white.}
		\label{fig:PIDConfMat}
\end{figure}

Figure~\ref{fig:PIDConfMat} summarizes the performance of the charge determination in the form of a confusion matrix. Each column shows the distribution of reconstructed nuclear charges, $Z_{\rm rec}$, for all events with true nuclear charge $Z_\mathrm{mc}$. The framework finds the true charge, or a very close one, for the majority of events. This suggests that the networks learn that nuclei with similar energy-deposition profiles have similar charges. Averaging over the full charge range, we find that 59\% of events are assigned to their true $Z$ class.

As expected, the accuracy, i.e., the fraction of correctly identified events for a given $Z_\mathrm{mc}$, decreases with increasing $Z$. Hydrogen (protons) and helium events are correctly identified in 99.8\% and 99.3\% of cases, respectively. For light nuclei up to oxygen, we achieve values well over 95\%. It is only for heavier nuclei that the exact charge determination evidently becomes more difficult. However, the network still learns to roughly recognize the energy-deposition profiles of nuclei with larger $Z$ and in most cases assigns the biggest fraction of events to the correct class.  Incorrectly reconstructed events are assigned to neighboring classes, with $Z_{\rm rec}$ close to $Z_\mathrm{mc}$.   

\begin{figure}
\centering
		\includegraphics[width=.9\linewidth]{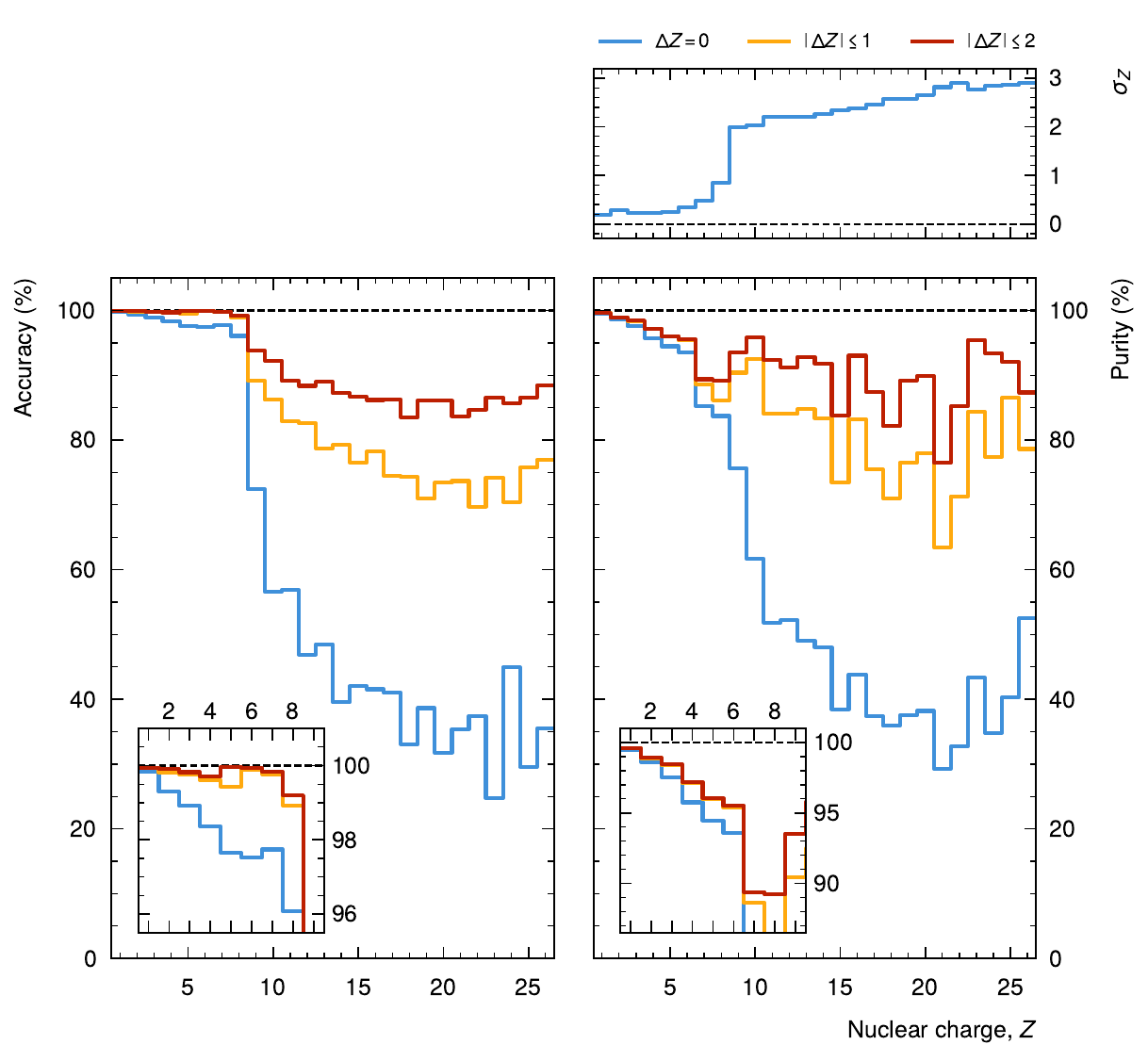}
		\caption{The accuracy (left panel) and purity (bottom right panel) of the charge determination as a function of $Z$. We define the accuracy as the fraction of correctly identified events for a given $Z_\mathrm{mc}$ and the purity as the fraction of events in a $Z_{\rm rec}$ class for which $Z_{\rm rec} = Z_{\rm mc}$. The top right panel shows the standard deviation, $\sigma_Z$, of the reconstructed-charge distributions (the columns in Fig.~\ref{fig:PIDConfMat}). The blue curves show the performance for an exact identification of the charge ($\Delta Z = Z_{\rm rec} - Z_{\rm mc} = 0$). The yellow and red curves illustrate the improvement when we allow $\left|\Delta Z\right| \le 1$ and $\left|\Delta Z\right| \le 2$, respectively. The insets in the lower panels provide zoomed-in views for $Z \le 8$.}
		\label{fig:PIDAcc}
\end{figure}

This is illustrated in Figure~\ref{fig:PIDAcc}, which shows the accuracy and the standard deviation, $\sigma_Z$, (here calculated as the square root of the variance) of the reconstructed-charge distributions as a function of $Z$. The blue curves show the performance for an exact identification of the charge ($\Delta Z = Z_{\rm rec} - Z_{\rm mc} = 0$). The fraction of correctly classified events slowly decreases from nearly 100\% (hydrogen) to a little over 96\% (oxygen), then drops to values below 75\% (fluorine) when the second network takes over. It ultimately reaches values between 30\% and 40\% for the highest $Z$. The same behavior is reflected in the width of the reconstructed-$Z$ distributions: For light elements, $\sigma_Z$ is close to 0.2; it then jumps to two for fluorine and steadily rises to almost three for iron. If we allow $\left|\Delta Z\right| \le 1$ (i.e., we count every event with $Z_{\rm mc} - 1 \le Z_{\rm rec} \le Z_{\rm mc} + 1$ as correctly identified), the values for light nuclei increase to over 99\% all the way to oxygen (yellow curve in Fig.~\ref{fig:PIDAcc}). Though we still observe a substantial break at the boundary between the two networks, the performance for heavy nuclei improves significantly and stays above 70\% for all elements. If we allow $\left|\Delta Z\right| \le 2$ (red curve), we obtain a further substantial improvement for large $Z$, with values consistently above 83\% across the whole range. In these cases, 84\% and 91\% of all events are assigned to their true $Z$ class, respectively. This behavior strongly suggests that our framework is able to reliably determine that a particle is, for example, iron-like even if it cannot exactly determine its charge.

We also investigated the purity of the charge determination, which we here define as the fraction of events in a $Z_{\rm rec}$ class for which $Z_{\rm rec} = Z_{\rm mc}$. It is 99.6\% and 98.8\% for hydrogen and helium, respectively, and exceeds 84\% for elements through oxygen. For heavier nuclei, it decreases to values around 35\%. For $\Delta Z = 0$, the mean purity over the entire charge range is 58\%. If we allow $\left|\Delta Z\right| \le 1$ or $\left|\Delta Z\right| \le 2$, the effect is analogous to that discussed above for the reconstruction accuracy (see bottom right panel of Fig.~\ref{fig:PIDAcc}), and the overall mean purity increases to 83\% and 91\%, respectively.

\subsection{Discussion}
The accuracy and purity of the charge determination for hydrogen (99.8\% and 99.6\%, respectively) and helium (99.3\% and 98.8\%) are fully compatible with the requirements of radiation monitoring. Together, these elements make up 99\% of all cosmic-ray nuclei, and the accuracy and purity of their identification is therefore crucial to dosimetry. For light nuclei ($Z \leq 8$), the performance for $\Delta Z = 0$ (accuracy and purity over 95\% and 84\%, respectively) is also adequate. The fact that accuracies better than 83\% for heavier elements can only be obtained if we allow $\left|\Delta Z\right| \le 2$ is not critical in the context of our work. Though the biological effectiveness of nuclei with similar charge differs somewhat due to the energy deposition's quadratic dependence on $Z$, the uncertainty introduced by allowing $\left|\Delta Z\right| \le 2$ is still much smaller than for a determination of $Z$ from the \gls{let} alone. Despite the reduced performance for heavy nuclei, the \gls{adu} can therefore measure the biologically relevant dose with higher accuracy than the majority of sensors used today.

We observe a sharp drop in performance at the interface between the low-$Z$ and the high-$Z$ network (see Fig.~\ref{fig:PIDAcc}). This suggests that the network architecture struggles most with learning to reconstruct the highest $Z$, which negatively affects the performance over the whole charge range the network is trained for. This hypothesis was corroborated by our attempts to shift the hand-over point between the networks to higher $Z$, which increased the accuracy of the low-$Z$ network for medium-light elements at the expense of a significantly reduced performance for hydrogen and helium. The difficulty in determining the charge of heavy nuclei reveals the inherently limited separation power of the detector: Ionization quenching in the plastic scintillators leads to ever smaller differences between the mean energy deposition of nuclei with larger $Z$ \citep{Losekamm2024}. At the same time, stronger energy-loss straggling and a higher fragmentation probability lead to larger variations in the energy-deposition profiles. Combined, these effects result in an increasingly large overlap in the profiles for particles with similar charge, restricting our ability to accurately determine $Z$ on an event-by-event basis. 

The impact of the network's difficulty to separate heavier nuclei on its performance for lighter ones also reveals one of the challenges of using machine-learning algorithms. During training, the network learns that the energy-deposition profiles of nuclei with higher $Z$ can vary widely. Since it attempts to find features that are common to all events, it (incorrectly) applies this knowledge to lighter elements. In the extreme---if the charge range of a network is chosen too wide---the bias introduced by large $Z$ can completely degrade the excellent performance that can be achieved for the clearly separable light elements. Therefore, a promising approach to improve the charge determination for medium-$Z$ elements might be to increase the number of networks, thereby reducing the charge range covered by each of them. A framework with more branches does, however, have a higher complexity (and thus longer execution time) and more hand-over points at which boundary effects can occur. In addition, a network architecture with more parameters may also help to increase the accuracy.

\section{Energy Measurement}
Determining the kinetic energy of nuclei traversing the \gls{adu} requires knowledge of their mass, which we determine from $Z$ by assuming a corresponding average $A$. We then use 26 element-specific neural networks of identical architecture (see Fig.~\ref{fig:NN_architectures}) to determine their energy. For low-energy nuclei that stop in the detector, the energy is given by the sum of their energy depositions. For through-going particles, the networks must learn to relate the energy-deposition profile to their velocity. They perform a regression analysis and return the energy per nucleon as a real number in the range of \SI{20}{MeV/n} to \SI{1}{GeV/n} in their single output node (except for the heaviest elements, for which the upper limit is \SI{10}{GeV/n}). We trained each network on an individual data set of 1.8 million simulated events and evaluated the combined framework using a separate data set of eight million events.

The purity of the subsets of events passed on to the energy-reconstruction networks ranges from 99.6\% for hydrogen to 52.3\% for iron (see previous section). At the same time, the error of reconstructing a nucleus' energy assuming a charge that is slightly off becomes smaller for heavier nuclei because of the decreasing relative difference in mass of adjacent elements. Therefore, to reduce the impact of an incorrectly reconstructed $Z$ for elements with $Z \ge 2$, the individual training data sets comprised events with charges in the range $[Z-1, Z+1]$ or $[Z-2, Z+2]$ (which is equivalent to allowing $\left|\Delta Z\right| \le 1$ or $\left|\Delta Z\right| \le 2$, see previous section), with all three (five) possible values equally likely.

\subsection{Results}
\begin{figure}[t]
\centering
		\includegraphics[width=.9\linewidth]{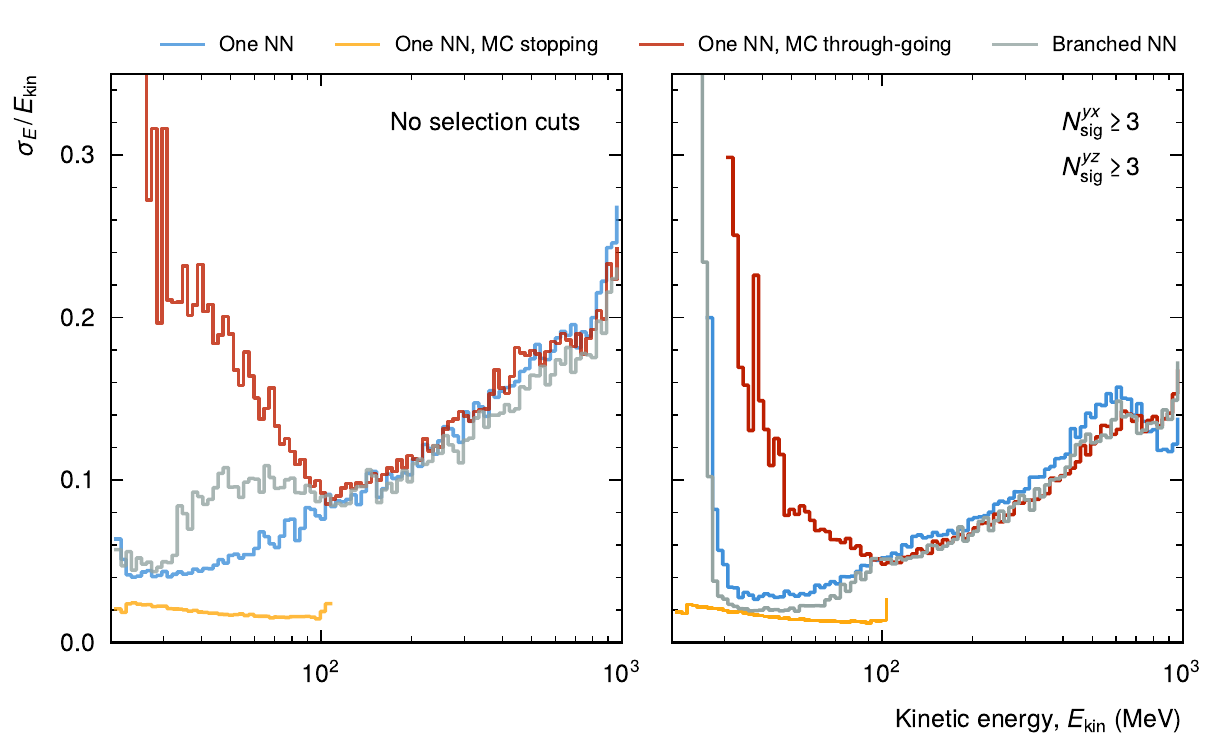}
		\caption{Bin-wise energy resolution for protons (hydrogen) with energies from \SI{20}{MeV} to \SI{1}{GeV}. The left panel shows the results if we apply no selection cuts to our test data, the right panel those for events with $N_{\rm sig}^{yx} \geq 3$ and $N_{\rm sig}^{yz} \geq 3$. The blue histograms give the performance of a single network trained over the full data set of stopping and through-going hydrogen nuclei; the yellow and red histograms that of separate networks for stopping and through-going particles. The gray histograms show the performance of a branched framework, where a filter network separates stopping from through-going particles and passes them to the separate energy-reconstruction networks.}
	\label{fig:H_Energy_Standard_Deviation_StoppingNotStopping}
\end{figure}

Measuring the energy of stopping nuclei in principle does not require a minimum number of fibers with a signal. To find the ideal lower limit of our detector's sensitivity range, we investigated two scenarios: one in which we applied no selection cuts to our test data set and one where we selected only events with $N_{\rm sig}^{yx} \geq 3$ and $N_{\rm sig}^{yz} \geq 3$ (like for charge determination). The results for hydrogen are illustrated in the left and right panels of Figure~\ref{fig:H_Energy_Standard_Deviation_StoppingNotStopping}, respectively. We show the bin-wise energy resolution, $\sigma_{E} / E_{\rm kin}$, where $\sigma_{\rm E}$ is the Gaussian width of the reconstructed energies for nuclei within a given energy bin and $E_{\rm kin}$ is the central value of that bin. For this purpose, we divide the respective energy ranges into 100 log-uniform bins.

The blue histograms show the resolution that a single network trained over the full data set of stopping and through-going hydrogen nuclei achieves. If we apply no selection cuts (see left panel), we obtain $\sigma_E/E_{\rm kin}\leq8\%$ for energies up to \SI{100}{\MeV}, for which most protons stop and deposit their full kinetic energy in the detector. The resolution monotonically decreases with increasing energies and reaches about 25\% at \SI{1}{\GeV}. For events with $N_{\rm sig}^{yx} \geq 3$ and $N_{\rm sig}^{yz} \geq 3$ (see right panel), we achieve a resolution of better than 5\% for $E_{\rm kin} \leq \SI{100}{\MeV}$ and better than 16\% for higher energies. Despite this significant overall improvement, we observe a worsening of the resolution at the lower end of the reconstruction range and a non-monotonic behavior towards the upper boundary. While the former effect is simply due to the small range of protons of such low energies and the correspondingly small number of events making it through the selection cuts, the reason for the latter one is not immediately clear.

We separately trained and tested our network on stopping and through-going particles (selected based on knowledge from the MC simulation) to assess how it copes with the different information content of the respective events. The results are shown in Figure~\ref{fig:H_Energy_Standard_Deviation_StoppingNotStopping} as yellow and red histograms. For stopping particles, we achieve $\sigma_E/E_{\rm kin}\leq2\%$ in both scenarios, with essentially no discernible difference up to \SI{100}{MeV}, above which the range of protons is larger than the size of the detector. For through-going particles, the resolution closely resembles that of the single-network reconstruction above \SI{100}{MeV} in the full data set and is slightly better if we apply selection cuts. The non-monotonic behavior at higher energies in the latter scenario is less pronounced. At lower energies, $\sigma_E/E_{\rm kin}$ substantially worsens with decreasing energy in either scenario, reflecting both the limited number of events in that range and the fact that through-going particles with such low energies traverse only few fibers.

We also investigated a more realistic scenario in which a neural network separates stopping from through-going particles and passes them to the respective energy-reconstruction network (see gray histograms in Fig.~\ref{fig:H_Energy_Standard_Deviation_StoppingNotStopping}). If we apply no selection cuts, this branched framework performs marginally better than the previously discussed networks above \SI{100}{MeV} but worse than the single-network reconstruction at lower energies. This suggests that the negative impact of incorrectly separating stopping from through-going particles outweighs the advantage of using a dedicated network for stopping ones. The framework's performance significantly improves if we require $N_{\rm sig}^{yx} \geq 3$ and $N_{\rm sig}^{yz} \geq 3$. At energies above \SI{100}{MeV}, the resolution is essentially identical to the case where we manually select through-going particles. Below \SI{100}{MeV}, it is noticeably better than that of the single-network reconstruction and approaches the lower limit of manually selected stopping protons for $E_{\rm kin} \leq \SI{50}{MeV}$. This indicates that our filter network can more effectively separate stopping from through-going particles due to the (on average) larger number of fiber signals per event. The improvement comes at the expense of sensitivity at the lowest energies, where $\sigma_E/E_{\rm kin}$ drastically worsens for $E_{\rm kin} \leq \SI{30}{MeV}$ because only few events pass the selection criteria.

\begin{figure}[t]
\centering
		\includegraphics[width=.9\linewidth]{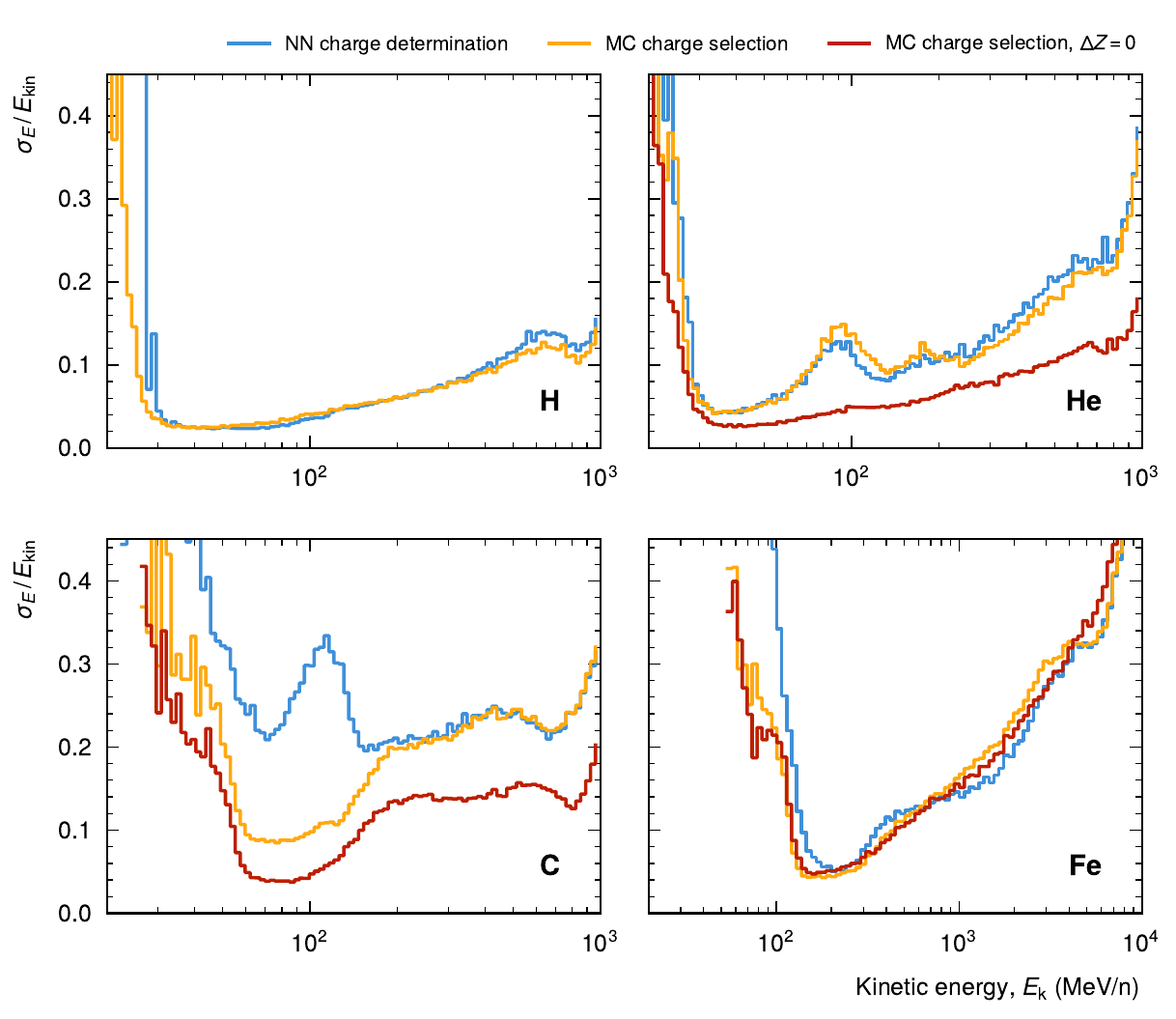}
		\caption{Energy resolution for hydrogen (H), helium (He), carbon (C), and iron (Fe) nuclei with energies from \SI{20}{MeV/n} to \SI{1}{GeV/n} (\SI{10}{GeV/n} for iron). The yellow histograms show the performance for events of the target charge which we manually selected, the blue ones the result for particles that were selected by our charge-determination network from a test data set containing all elements up to cobalt. For helium and carbon, we trained the networks with nuclei in the range $[Z-1, Z+1]$ of the target charge and the one for iron in the range $Z\in[24,28]$. The red histograms show the performance for manually selected events for networks trained only with the target charge. The network for hydrogen is always trained for the target charge only.}
		\label{fig:Energy_Standard_Deviation}
\end{figure}

Finally, we evaluated the performance of the single-network approach for heavier nuclei. We used a data set comprising all elements up to cobalt for this analysis. Figure~\ref{fig:Energy_Standard_Deviation} exemplarily summarizes the results for hydrogen (H), helium (He), carbon (C), and iron (Fe) nuclei. It shows the energy resolution for events with manually selected $Z$ (yellow histograms) and for particles that were assigned to a given $Z$ class by our charge-determination network (blue histograms). In both cases, we trained the networks for He and C with nuclei in the range $[Z-1, Z+1]$ of the target charge and the one for Fe ($Z = 26$) with nuclei in the range $Z\in[24, 28]$. For comparison, the red histograms show the ideal performance, for which we trained all networks on the target charge only (i.e., $\Delta Z = 0$) and manually selected events based on knowledge from the \gls{mc} simulation. The network for hydrogen is always trained for the target charge only.

For hydrogen, the blue and yellow histograms allow to investigate the impact of charge confusion (i.e., data sets with sub-100\% purities being passed on to the energy-measurement networks). We achieve the previously discussed good resolution in the case of manually selected events, and we see that wrongly assigned nuclei (primarily helium) affect it below \SI{30}{MeV} and above about \SI{200}{MeV}. The large decrease in resolution at the low-energy bound is intuitively understandable because $\Delta Z \ge 1$ has a huge impact on the energy loss of stopping light nuclei. The effect can be reduced by training the network for charges in the range $Z\in[1,2]$. This, however, comes at the expense of a lower resolution across all energies, which has a larger impact on the overall performance than improving the lower sensitivity limit and is therefore not desirable. For through-going particles, the maximum difference is about two percentage points and occurs at around \SI{600}{MeV}. In either case, we achieve $\sigma_E/E_{\rm kin}\leq5\%$ for $E_{\rm kin} < \SI{100}{MeV}$, $\sigma_E/E_{\rm kin}\leq10\%$ for $E_{\rm kin} < \SI{300}{MeV}$, and $\sigma_E/E_{\rm kin}\leq16\%$ for $E_{\rm kin} < \SI{1}{GeV}$.

For helium, the difference between the blue and yellow histograms is less pronounced. This illustrates that the effect of charge confusion can largely be suppressed if the network sees neighboring elements during training. The red histogram (training with $\Delta Z = 0$ and evaluation with manually selected helium events), shows that in the unrealistic case of perfect charge determination, we get a performance comparable to that for hydrogen. The largest differences are a ten-percentage-point bump around \SI{90}{MeV/n} and an even larger divergence toward the upper energy boundary. Overall, we achieve $\sigma_E/E_{\rm kin}\leq14\%$ for $E_{\rm kin} < \SI{300}{MeV/n}$ and $\sigma_E/E_{\rm kin}\leq24\%$ for $E_{\rm kin} < \SI{800}{MeV/n}$.

The case of carbon is exemplary for medium-$Z$ nuclei and shows that charge confusion can affect the energy resolution much more significantly than observed for helium. The impact is most pronounced at around \SI{100}{MeV/n}, where the blue histogram for neural-network-based charge determination peaks at almost 34\%. The yellow curve for manually selected events, on the other hand, is at around 10\%, only about four percentage points above the ideal limit (red histogram). At higher energies, the blue and yellow curves agree almost perfectly, indicating that we need to further look into (and understand) the energy dependence of the charge confusion. Realistically achievable $\sigma_E/E_{\rm kin}$ for carbon appear to be in the in 10\%-to-25\% range.

The results for iron are overall better than for carbon, and the curves for the three evaluations largely agree. The fully neural-network-based framework (blue histogram) outperforms the supposedly ideal case (red histogram) for some energies above \SI{1}{GeV/n}. Here, it seems that the combined effects of charge confusion and the broader training set ($|\Delta Z| \leq 2$) more than compensate for the uncertainty introduced by the low purity of the data set selected by the charge-determination network. The exact cause, however, is not yet clear and requires further investigation. Overall, we achieve $\sigma_E/E_{\rm kin}\leq7\%$ for $E_{\rm kin} \approx \SI{150}{MeV/n}$, $\sigma_E/E_{\rm kin}\leq10\%$ for $E_{\rm kin} < \SI{400}{MeV/n}$, and $\sigma_E/E_{\rm kin}\leq20\%$ for $E_{\rm kin} < \SI{2}{GeV/n}$.

\subsection{Discussion}
Our results show that energy resolutions on the order of 10\% to 20\% can be achieved, with even better values for hydrogen and helium over a wide range of the most relevant particle energies. However, the cases of helium, carbon, and iron also demonstrate that our charge measurement can still be improved. There are many features whose cause we do not yet fully understand but which suggest that both the charge-determination and energy-measurement networks face similar challenges. The present results nonetheless demonstrate that an energy measurement based on a neural-network-based reconstruction is possible and can deliver acceptable resolutions. An important aspect of our future work is to find a better approach for dealing with charge confusion and minimizing its impact on the energy measurement. We also need to introduce a mechanism that allows networks to deal with particle energies outside their target range.

\section{Conclusion}
In summary, our findings show that a detailed characterization of cosmic-ray nuclei with energies in the \si{MeV}-to-\si{GeV} range can be performed with a simple and compact instrument like the RadMap Telescope's \gls{adu}. They also reveal the inherent physical limits of the detector and highlight some of the challenges of using a reconstruction framework based on neural networks. The results presented here are based on a simulation that is largely realistic yet does not take into account some of the aspects of a real detector that are harder to model, for example optical and electrical crosstalk, and the misalignment of individual scintillating fibers. The study of how well our network architecture can cope with such effects remains open work.

We already discussed the performance and limits of the networks for the individual reconstruction tasks and how they might be improved. Now, we also want to draw attention to potential improvements that are common to them. 

For example, enhancements in all three tasks can surely be achieved by refining the training of our networks. For example, we deliberately used training data in which all elements are equally abundant to avoid biases towards those with a naturally higher flux. We also did so because using realistic abundances is impractical, as the fluxes of the most and least abundant elements differ by almost six orders of magnitude. The knowledge that the chances of encountering certain elements in the real cosmic-ray environment may differ significantly will very likely improve our framework's overall performance. A similar argument can be made for the energy spectrum, which we likewise did not model realistically to reduce the computational effort of training. We therefore need to develop an approach for teaching our networks the real elemental and spectral abundances without using fully realistic training data. The challenge of doing so lies in ensuring that the networks are optimally trained across their whole sensitivity range.

Some of the framework's limitations may be caused by our networks not being able to capture all information contained in the detector's events.
We chose rather simple convolutional architectures that make use of a representation of event signatures as two images. A multitude of more complex approaches has since been applied to track and characterize particles in detectors. It was shown that transformer architectures can perform a variety of tasks \citep{Vaswani2017}; graph neural networks are now widely used to reconstruct particle tracks \citep{Scarselli2009}; and physics-informed networks may be useful for refining our charge and energy determination \citep{Raissi2019}. Part of our future work is to investigate the impact of a more complex architecture on our reconstruction performance and computational requirements.

Another limiting factor common to all reconstruction tasks is the little information contained in events with only few fiber signals, for example of particles that have very low energies or traverse one of the corners of the detector. Tracks or energy-deposition profiles with fewer data points suffer more from statistical fluctuations caused by straggling and path-length differences than longer ones. Though we have not yet studied this effect systematically, the case of the track reconstruction shows that such events can have substantially higher uncertainties. Excluding them from analysis, potentially at the expense of sensitivity at the lowest energies, may improve the overall performance of the networks, though care must be taken to avoid introducing biases during event selection.

Finally, all our results are based on simulations for a detector in open space, i.e., with almost no material around it. This is, of course, not a realistic application scenario. In the RadMap Telescope, the \gls{adu} is contained in a housing made from aluminum and surrounded by layers of electronics with a non-uniform material budget. The instrument is deployed inside the \gls{iss}, which provides yet more substantial shielding. Even in a (hypothetical) scenario where the detector was mounted to the outside of a spacecraft with little need for protection against the space environment, significant shielding across parts of the solid angle would be provided by the spacecraft itself. However, neglecting the effects from additional material allowed us to evaluate the capabilities of the instrument and our reconstruction framework under near-optimal conditions. In the context of our work, the positive effect of shielding is that the relative contribution of hydrogen and helium to the absorbed dose becomes ever more dominant for increasing material thickness \citep{Slaba2017}. Since our charge determination and energy measurement work best for these light nuclei, this leads to a smaller uncertainty of the biologically relevant dose calculated from our measurements.
 
In conclusion, our study demonstrates that the RadMap Telescope can perform spectroscopic measurements that significantly exceed the capabilities of most instruments used for radiation monitoring aboard spacecraft today. It also shows that using a neural network-based reconstruction is feasible and can produce acceptable results. Even though the numerical values cited here may shift slightly (for better or worse) once we implemented a more realistic detector model and changes to our framework, we very much expect this qualitative conclusion to hold.

\begin{acknowledgements}
Our work was funded by the Deutsche Forschungsgemeinschaft (DFG, German Research Foundation) via grant 414049180 and via Germany's Excellence Initiative -- EXC-2094 -- 390783311.
\end{acknowledgements}

\bibliography{NNpaper}
   

\end{document}